\documentclass[reprint,superscriptaddress,aps,prl,floatfix]{revtex4-1}

\usepackage{dcolumn}% Align table columns on decimal point
\usepackage{bm}% bold math
\usepackage{amsmath, color, amssymb, latexsym, graphicx, lipsum}
\newcommand{\ua}{\textbf{u}_{\text{A}}}
\newcommand{\da}{\boldsymbol\nabla\cdot\textbf{u}_{\text{A}}}
\newcommand{\dap}{\boldsymbol\nabla'\cdot\textbf{u}_{\text{A}}'}
\newcommand{\uh}{\textbf{u}}
\newcommand{\dv}{\boldsymbol\nabla\cdot\textbf{u}}
\newcommand{\dvp}{\boldsymbol\nabla'\cdot\textbf{u}'}
\newcommand{\ja}{\textbf{J}_{\text{c}}}
\newcommand{\dja}{\boldsymbol\nabla\cdot\textbf{J}_{\text{c}}}
\newcommand{\djap}{\boldsymbol\nabla'\cdot\textbf{J}_{\text{c}}'}
\def\ADD#1{{\textcolor{black}{#1}}}   % addition for authors
\def\ADDD#1{{\textcolor{black}{#1}}}   % addition for authors

\begin{document}

\preprint{APS/123-QED}

\title{Energy cascade rate measured in a collisionless space plasma with MMS data and compressible Hall magnetohydrodynamic turbulence theory}

\author{N. Andr\'es}
\affiliation{Laboratoire de Physique des Plasmas, \'Ecole Polytechnique, CNRS, Sorbonne University, Observatoire de Paris, Univ. Paris-Sud, F-91128 Palaiseau Cedex, France}
\affiliation{Instituto de Astronom\'ia y F\'{\i}sica del Espacio, CONICET-UBA, Ciudad Universitaria, 1428, Buenos Aires, Argentina}
\affiliation{Departamento de F\'{\i}sica, Facultad de Ciencias Exactas y Naturales, UBA, Ciudad Universitaria, 1428, Buenos Aires, Argentina}
\email{nandres@iafe.uba.ar, nandres@df.uba.ar}
\author{F. Sahraoui}
\affiliation{Laboratoire de Physique des Plasmas, \'Ecole Polytechnique, CNRS, Sorbonne University, Observatoire de Paris, Univ. Paris-Sud, F-91128 Palaiseau Cedex, France}
\author{S. Galtier}
\affiliation{Laboratoire de Physique des Plasmas, \'Ecole Polytechnique, CNRS, Sorbonne University, Observatoire de Paris, Univ. Paris-Sud, F-91128 Palaiseau Cedex, France}
\affiliation{Institut universitaire de France, France}
\author{L. Z. Hadid}
\affiliation{European Space Agency, ESTEC, Noordwijk, Netherlands}
\author{R. Ferrand}
\affiliation{Laboratoire de Physique des Plasmas, \'Ecole Polytechnique, CNRS, Sorbonne University, Observatoire de Paris, Univ. Paris-Sud, F-91128 Palaiseau Cedex, France}
\author{S.Y. Huang}
\affiliation{School of Electronic and Information, Wuhan University, Wuhan, China}
\affiliation{Laboratoire de Physique des Plasmas, \'Ecole Polytechnique, CNRS, Sorbonne University, Observatoire de Paris, Univ. Paris-Sud, F-91128 Palaiseau Cedex, France}

\date{\today}

\begin{abstract}
The first complete estimation of the compressible energy cascade rate $|\varepsilon_\text{C}|$ at magnetohydrodynamic (MHD) and sub-ion scales is obtained in the Earth's magnetosheath using Magnetospheric MultiScale (MMS) spacecraft data and an exact law derived recently for {\it compressible} Hall MHD turbulence. A multi-spacecraft technique is used to compute the velocity and magnetic gradients, and then all the correlation functions involved in the exact relation. It is shown that when the density fluctuations are relatively small, $|\varepsilon_\text{C}|$ identifies well with its incompressible analogue $|\varepsilon_\text{I}|$ at MHD scales but becomes much larger than $|\varepsilon_\text{I}|$ at sub-ion scales. For larger density fluctuations, $|\varepsilon_\text{C}|$ is larger than $|\varepsilon_\text{I}|$ at every scale with a value significantly higher than for smaller density fluctuations. Our study reveals also that for both small and large density fluctuations, the non-flux terms remain always negligible with respect to the flux terms and that the major contribution to $|\varepsilon_\text{C}|$ at sub-ion scales comes from the compressible Hall flux. 
\end{abstract}

\maketitle

%%%%%%%%%%%%%%%%%%%%%%%%%
{\it Introduction.} 
Turbulence is a universal phenomenon observed from quantum to astrophysical scales \cite[see e.g.,][]{Al2018,G2018,Po2018}. 
It is mainly characterized by a nonlinear transfer (or cascade) of energy from a source to a sink. In astrophysical plasmas, fully developed turbulence plays a major role in several physical processes such as accretion flows around massive objects, star formation, solar wind heating or energy transport in planetary magnetospheres \citep[e.g.,][]{B2005,S2009,Ch2015,H2015}. In particular, the Earth's magnetosheath (MS) -- the region of the solar wind downstream of the bow shock \citep{K1995} -- provides a unique laboratory to investigate {\it compressible} plasma turbulence. Indeed, a key feature of the MS plasma is the high level of density fluctuations in it, which can reach up to $100 \%$ of the background density \citep{S1992,S2003,Hu2017}, in contrast to the solar wind where it is $\sim 20\%$ \citep{H2012,H2017a}. Thanks to the high time resolution data provided by the Magnetospheric MultiScale (MMS) mission in the Earth's MS \citep{Bu2016}, we are now able for the first time to measure the compressible energy cascade rate $\varepsilon_\text{C}$ from the magnetohydrodynamic (MHD) inertial range down to the sub-ion scales. 

The energy cascade rates can be estimated using exact laws derived from fluid models which relate, in the simplest case, the longitudinal structure functions of the turbulent variables (e.g., the velocity field {\bf u} or the magnetic field {\bf B}) taken in two points, to the spatial increment $\boldsymbol{\ell}$ that separates them. The first exact relation for a plasma was derived by \citet{P1998a,P1998b} (hereafter PP98): it describes three-dimensional (3D) incompressible MHD (IMHD) turbulence under the assumption of statistical homogeneity and isotropy. This law has been the subject of several numerical tests \citep[see, e.g.][]{Mi2009,Bo2009,W2010}; it has been used for the evaluation of the incompressible cascade rates in space plasmas \citep{SV2007,Sa2008,Co2015} (denoted here $\varepsilon_\text{IMHD}$) and the magnetic/kinetic Reynolds numbers \citep{WEY2007}, and for the large-scale modeling of the solar wind \citep{M1999,Mc2008}. 

The IMHD approximation has been successfully used to study plasma turbulence at scales larger than the ion inertial length $d_i$ (or  the Larmor radius $\rho_i$) \citep[see, e.g.][]{M1982,B2005}. However, at spatial scales comparable or smaller than $d_i$, the ions are no longer frozen-in to the magnetic field lines because of the Hall term in the generalized Ohm's law \citep[e.g.,][]{K1995}. Moreover, the incompressibility assumption is likely to fail to describe sub-ion scales physics: it is theoretically justified at MHD scales because of the existence of a purely incompressible Alfv\'en wave solution. However, that mode becomes a Kinetic Alfv\'enic Wave (KAW) at sub-ion scales, which is inherently compressible since it carries density fluctuations \citep{S2007, S2009}. A nearly incompressible whistler mode can develop at high frequency \citep{B2012}, but that mode is unlikely to dominate the sub-ion scales cascade in the solar wind or MS \citep{S2009,P2012,K2013,D2019}. These considerations emphasize the crucial need to incorporate density fluctuations in the description of the sub-ion scale cascade, as we will show below using MMS data in the MS. 

Following the same methodology as in \citep{Ga2011}, \citet{B2013} derived an exact law for isothermal compressible MHD (CMHD) turbulence. Recently, \citet{A2017b} revisited that work by providing an alternative derivation of the exact law that relates the compressible energy cascade rate (hereafter $\varepsilon_\text{CMHD}$) to four different categories of terms, namely the source, hybrid and $\beta$-dependent terms, in addition to the well-known flux terms. Using the model of \citep{B2013} and in situ measurements from the THEMIS spacecraft~\citep{Au2009}, \citet{B2016c} and \citet{H2017a} evidenced the role of density fluctuations in amplifying the energy cascade rate in the slow wind compared to the fast wind.  \citet{H2017b} have further found that density fluctuations reinforce the anisotropy of the energy cascade rate with respect to the local magnetic field in the Earth's MS, and evidenced a link with kinetic plasma instabilities. However, those observational works were limited to the inertial range and used only some of the flux terms (all the source and the majority of the hybrid terms could not have been estimated using single spacecraft data). 

In this Letter, we provide the first complete estimation of the compressible turbulent energy cascade rate in the MHD inertial range {\it and} at the sub-ion scales ($\varepsilon_\text{CHall}$) in a collisionless plasma. We use the MMS high time resolution observations made in the Earth's MS and an exact relation recently derived for compressible Hall-MHD (HMHD) turbulence \citep{A2018}. We investigate the impact of the level of density fluctuations on $\varepsilon_\text{CMHD}$ and $\varepsilon_\text{CHall}$ by their comparison to $\varepsilon_\text{IMHD}$ and $\varepsilon_\text{IHall}$ obtained respectively with incompressible MHD and HMHD theories \citep{P1998a,P1998b,Ga2008,H2018,F2019}.

%%%%%%%%%%%%%%%%%%%%%%%%%
{\it Theoretical model.}
Using the compressible HMHD equations \citep[e.g.,][]{Ga2016} and following the usual assumptions for fully developed homogeneous turbulence (i.e., infinite kinetic and magnetic Reynolds numbers and a steady state with a balance between forcing and dissipation \citep{Ga2011,B2018}), an exact relation for fully developed turbulence can be obtained \citep{A2018} as,
\begin{widetext}
\ADD{
\begin{align}\nonumber
\varepsilon_\text{C} = &~
    \varepsilon_\text{CMHD}+\varepsilon_\text{CHall} = (\varepsilon^{flux}_\text{CMHD}+\varepsilon^{non-flux}_\text{CMHD})+(\varepsilon^{flux}_\text{CHall}+\varepsilon^{non-flux}_\text{CHall}) = \\ \nonumber
    & -\frac{1}{4}\boldsymbol\nabla_\ell\cdot\big\langle [(\delta(\rho\uh)\cdot\delta\uh+\delta(\rho\ua)\cdot\delta\ua + 2\delta e\delta\rho\big]\delta\uh - [\delta(\rho\uh)\cdot\delta\ua+\delta\uh\cdot\delta(\rho\ua)]\delta\ua \big\rangle \\ \nonumber
    & - \frac{1}{4}\langle\big(e_c'+\frac{u_\text{A}}{2}^{'2}\big)\boldsymbol\nabla\cdot(\rho\uh) + (e_c+\frac{u_\text{A}}{2}^2\big)\boldsymbol\nabla'\cdot(\rho'\uh')\rangle + \frac{1}{4}\langle\beta_i^{-1'}\boldsymbol\nabla'\cdot(e_c'\rho\uh) + \beta_i^{-1}\boldsymbol\nabla\cdot(e_c\rho'\uh') \rangle \\ \nonumber
    &-\frac{1}{2}\langle\big(R_E'-\frac{R_B'+R_B}{2}-E'+\frac{P_M'-P'}{2}\big)(\dv)+\big(R_E-\frac{R_B+R_B'}{2}-E+\frac{P_M-P}{2}\big)(\dvp)\rangle \\ \nonumber
    &-\frac{1}{2}\langle\big[R_H-R_H'-\bar{\rho}(\uh'\cdot\ua)+H'\big](\da)+\big[R_H'-R_H-\bar{\rho}(\uh\cdot\ua')+H\big](\dap)\rangle \\ \nonumber
    &-\frac{1}{2}\boldsymbol\nabla_\ell\cdot\big\langle [(\overline{\rho\ja\times\ua})\times\delta\ua-\delta(\ja\times\ua)\times\overline{\rho\ua}] \big\rangle \\  \label{exact_law}
    &-\frac{1}{2}\langle\delta\rho\frac{\ja\cdot\ua'}{2}(\da)-\delta\rho\frac{\ja'\cdot\ua}{2}(\dap)\rangle-\frac{1}{4}\langle(R_B-R_B')(\dja)+(R_B'-R_B)(\djap)\rangle,
\end{align}
}
\end{widetext}
with, by definition, $\rho$ the mass density, $\ua \equiv {\bf B}/\sqrt{4 \pi \rho}$ the compressible Alfv\'en velocity, $e_c \equiv c_s^{2} \log(\rho/\langle \rho\rangle)$ the internal energy, $c_s$ the local sound speed, $\ja \equiv {\bf J}/en$ the compressible electric current, ${\bf J} \equiv (c/4\pi) \nabla \times {\bf B}$ the current density, $n$ the ion number density, $e$ the electron charge, $P=c_s^2\rho$ the pressure, $P_M \equiv \rho u_{A}^2/2$ the magnetic pressure, $E \equiv \frac{\rho}{2}(\uh\cdot\uh+\ua\cdot\ua) + \rho e$ and  $H \equiv \rho(\uh\cdot\ua)$ are respectively the one-point total energy and density-weighted cross-helicity per unit volume, $R_E\equiv\frac{\rho}{2}(\uh\cdot\uh'+\ua\cdot\ua') + \rho e'$ and $R_H\equiv\frac{\rho}{2}(\uh\cdot\ua'+\ua\cdot\uh')$ are their respective two-point correlation functions, and $R_B\equiv\frac{\rho}{2}(\ua\cdot\ua')$  is magnetic energy density. Fields are taken at point $\textbf{x}$ or $\textbf{x}'=\textbf{x}+\boldsymbol\ell$; in the latter case a prime is added to the field. The angular bracket $\langle\cdot\rangle$ denotes an ensemble average \citep{Ba1953}, which is taken here as time average assuming ergodicity. We have introduced the usual increments and local mean definitions, i.e., $\delta\alpha\equiv\alpha'-\alpha$ and $\bar{\alpha}\equiv(\alpha'+\alpha)/2$ (with $\alpha$ any scalar {or vector} function), respectively. 

\ADD{From Eq.~\eqref{exact_law}, one can notice that the compressible energy cascade rate $\varepsilon_\text{C}$ can be split into two components: a purely MHD component $\varepsilon_\text{CMHD}$ (2nd to 5th lines) \citep[][hereafter, AS17]{A2017b} and a sub-ion one $\varepsilon_\text{CHall}$ (6th to 7th lines) \citep[][hereafter, AGS18]{A2018}, which corresponds to the terms in Eq.~\eqref{exact_law} proportional to $\ja$. In addition, for each component, the cascade can be split into two types of terms: a flux term that can be written as the local divergence of products of two-point increments and {\it non-flux} terms that involve spatial divergence of the different fields (e.g., $\uh$, $\ua$ or $\rho\uh$). The flux terms are the usual terms present in exact laws of incompressible  turbulence \citep{P1998a,P1998b}. Assuming isotropy, these flux terms reflect the nonlinear cascade rate of energy through scales, while the non flux terms act on the inertial range as a source or a sink for the mean energy cascade rate \citep[see,][]{Ga2011,A2017b}}

\begin{figure}
  \centering
  \includegraphics[width=.45\textwidth]{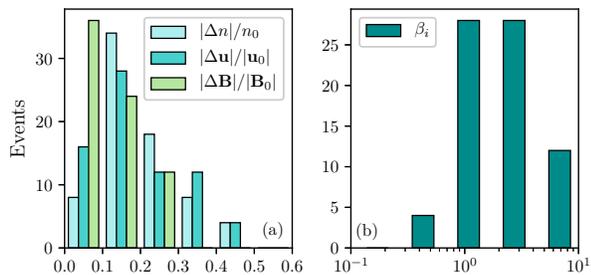}
  \caption{\ADD{Histogram for (a) the number density, velocity and magnetic fluctuations and (b) the ion $\beta_i$ parameter, respectively.}}
\label{histos}
\end{figure}

Expression \eqref{exact_law} is an exact relation describing homogeneous, stationary, isothermal, compressible HMHD turbulence \citep{A2018}. It generalizes previous laws \citep{P1998a,P1998b,Ga2011,B2013,A2017b} by including plasma compressibility, spatial anisotropy and the Hall effect. On the one hand, when incompressibility is assumed, Eq.~\eqref{exact_law} reduces to the exact law for incompressible HMHD turbulence with $\varepsilon_\text{I} = \varepsilon_\text{IMHD} + \varepsilon_\text{IHall}$, where $\varepsilon_\text{IHall}$ is the contribution of the Hall term \citep[][hereafter, F19]{Ga2008,F2019}. On the other hand, for scales larger than $d_i$ (MHD limit), the terms proportional to $\ja$ go to zero and Eq.~\eqref{exact_law} converges toward the exact law of CMHD turbulence \citep{A2017b,B2013}.

\begin{figure}
  \centering
  \includegraphics[width=.49\textwidth]{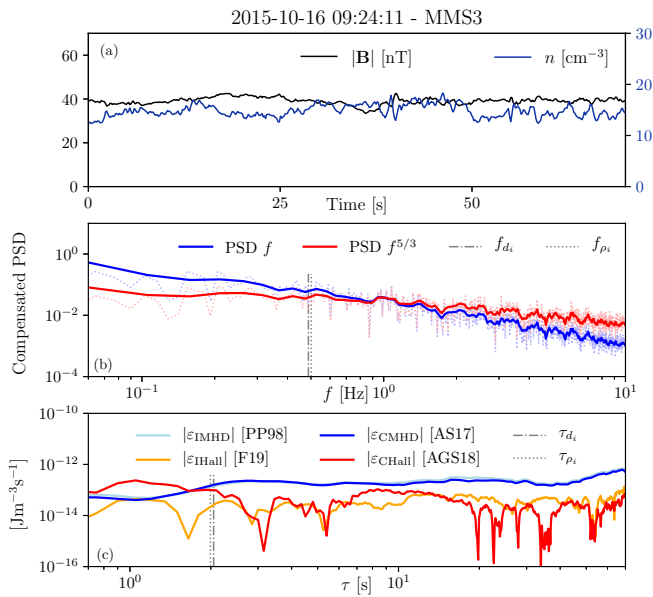}
  \caption{\ADD{(a) Magnetic field amplitude (black) and number density (blue), as a function of time.~(b) Magnetic field spectra compensated by the power laws $f$ (blue) and $f^{5/3}$ (red), as a function of the frequency $f$ \ADDD{(solid bold line is the average spectrum obtained using a sliding window)}.~(c) Energy cascade rates estimated from incompressible MHD (light blue), incompressible HMHD (orange), compressible MHD (blue) and compressible HMHD (red), as a function of the time lag $\tau$. Vertical dash-dot in and dot gray lines in (b) and (c) correspond to the Taylor shifted ion skin depth and Larmor radius.}}
\label{fig:1}
\end{figure}

\begin{figure}
  \centering
  \includegraphics[width=.49\textwidth]{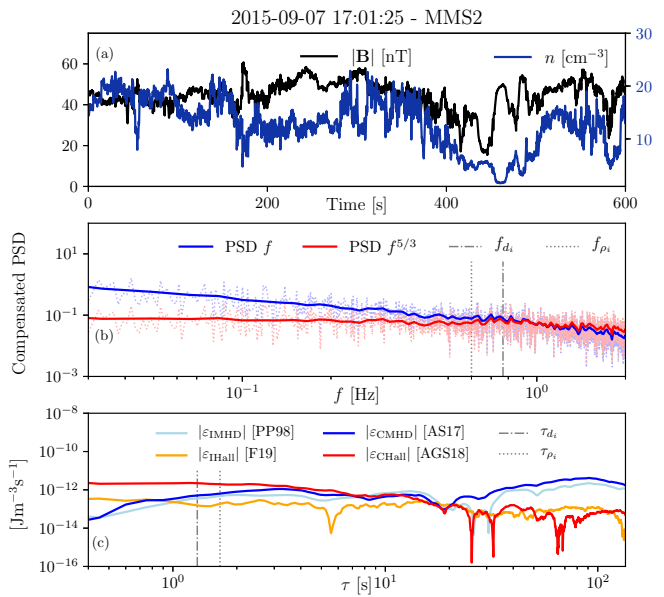}
  \caption{Same plots as in Fig.~\ref{fig:1} but for another time interval.}
\label{fig:2}
\end{figure}

%%%%%%%%%%%%%%%%%%%%%%%%%
{\it MMS data selection.} 
To compute each term in the RHS of Eq. \eqref{exact_law}, we used MMS spacecraft data in burst mode \citep{Bu2016} and during intervals of time when it was traveling in the Earth's MS. The magnetic field data and ion plasma moments were measured respectively by the Flux Gate Magnetometer (FGM) \citep{R2016} and the Fast Plasma Investigation (FPI) Dual Ion/Electron Sensors (DIS/DES) \citep{P2016}. The data sampling time is 150ms, set by the lowest sampling rate, i.e. that of the FPI Ion Sensor \citep{R2016}.

When we use the Taylor hypothesis on spacecraft measurements, the time sampling of the data is converted into a one-dimensional spatial sampling of the turbulent fluctuations along the flow direction \citep{B2016c,H2017a,P2018}. Therefore, we had constructed temporal correlation functions of the different turbulent fields at different time lags $\tau$ taken within the interval $\sim$ [$0.15$ -- $300$] s, which allow us to probe into MHD and sub-ion scales. The terms that include divergence of the fields in Eq.~\eqref{exact_law} (i.e., the source, hybrid and $\beta$-dependent terms) \citep[see,][]{A2017b,A2018} involve spatial derivatives that were fully computed using the four multi-spacecraft data of MMS \citep{P2000}. The electric current comes from the ion and electron moments measured by FPI. The results on the cascade rate were checked against the current estimates given by the curlometer technique \citep{D1988}, and no significant difference was found. 

\ADD{In a large survey of the Cluster data in the Earth's MS, \citet{Hu2017} found that the magnetic field fluctuations at the MHD scales near the bow shock have generally a power spectral density (PSD) close to $f^{-1}$ \citep[see also,][]{M2018}, whose physics is still largely unknown, while those that have a Kolmogorov-like spectrum (i.e., $f^{-5/3}$) were observed toward the flanks of the magnetopause. Since this study focuses on turbulence cascade and uses theoretical models that assume the existence of the inertial range, we selected only cases that showed a Kolmogorov-like spectrum in the MHD scales. However, some spectra showed small bumps near the ion scale, which would reflect the presence of kinetic instabilities \citep{S2006}. This aspect is not investigated in this first work on sub-ion scale cascade. Our data selection criteria resulted in a total of 72 intervals of $\sim$ 300 s each. In particular, Fig.~\eqref{histos} shows the histograms for (a) the number density, velocity and magnetic field fluctuations and (b) the ion $\beta_i$ parameter, respectively.}

\begin{figure}
  \centering
  \includegraphics[width=.45\textwidth]{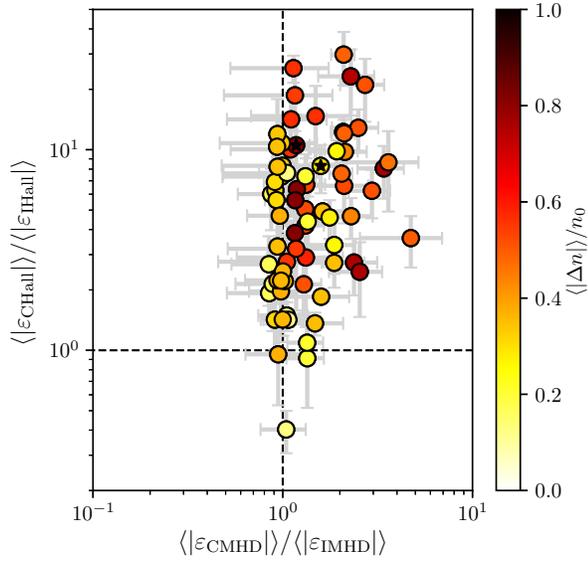}
  \caption{Mean ratio of the compressible to incompressible cascade rates $|\varepsilon_\text{CHall}|/|\varepsilon_\text{IHall}|$ as function of the mean ratio $|\varepsilon_\text{CMHD}|/|\varepsilon_\text{IMHD}|$. The color bar indicates the mean relative density fluctuations per event (72 events are analyzed) in the MS plasma. \ADDD{The error bars are the corresponding standard deviation in the MHD and sub-ion ranges.} The stars correspond to cases in Figs \ref{fig:1} and \ref{fig:2}.}
\label{fig:4}
\end{figure}

\begin{figure}
  \centering
  \includegraphics[width=.45\textwidth]{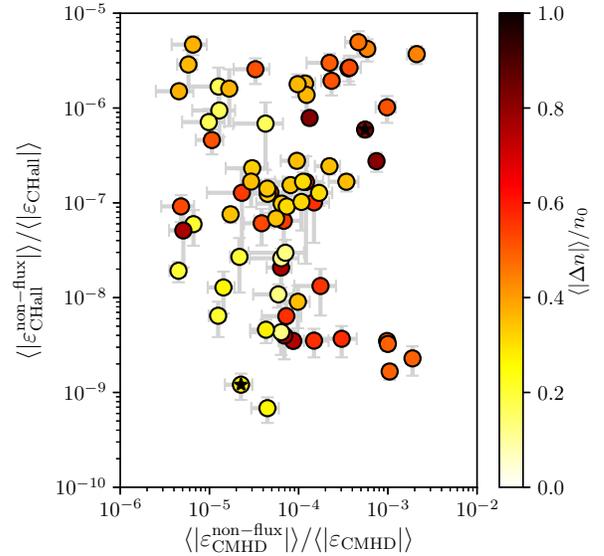}
  \caption{Mean ratio of the non-flux terms to the compressible cascade rates in the sub-ion scales $|\varepsilon^\text{non-flux}_\text{CHall}|/|\varepsilon_\text{CHall}|$ as function of ratio in the MHD scales $|\varepsilon^\text{non-flux}_\text{CMHD}|/|\varepsilon_\text{IMHD}|$. The color bar indicates the mean level of density fluctuations per event (72 events are analyzed) in the MS plasma. \ADDD{The error bars are the corresponding standard deviation in the MHD and sub-ion ranges.} The stars correspond to cases in Figs \ref{fig:1} and \ref{fig:2}.}
\label{fig:3}
\end{figure}

%%%%%%%%%%%%%%%%%%%%%%%%%
{\it Results.} 
For the selected events, we computed the total energy cascade rates $\varepsilon_\text{C}$ using the exact relation \eqref{exact_law} \citep{A2018}. Figs. \ref{fig:1} and \ref{fig:2} (c) show two representative examples of both MHD and sub-ion scales contributions to the total energy cascade rate from the incompressible and compressible exact relations. We emphasize here that we are only considering the magnitude of the cascade rate rather than its signed value. The latter requires much larger statistical samples to ensure statistical convergence \citep{H2017b, Co2015}, which are not yet available to us due to the limited duration of the burst mode on MMS \citep[see,][]{Bu2016}. \ADD{Signed cascade rates are relevant to study the direction (i.e., direct vs. inverse) of the energy cascade, which is beyond the scope of this Letter.} 

Figures \ref{fig:1}, \ref{fig:2} (a) and (b) show respectively the magnetic field amplitude and number density, and the compensated PSD for the magnetic energy. While both examples show a Kolmogorov-like slope at the largest scales, the level of density fluctuations and its correlation with the magnitude of $|{\bf B}|$ are very different. In Fig.~\ref{fig:1} (a), we observe a relatively uniform $|{\bf B}|$, a clear feature of incompressible Alfv\'enic fluctuations \citep[e.g., ][]{K2009}, in agreement with the finding of \citet{C2017} who analyzed this event: they found the dominance of Alv\'enic turbulence at the MHD scales which transitions into KAW near the ion scales. In contrast, Fig.~\ref{fig:2} (a) shows higher fluctuations in the amplitude of the magnetic field, which are anti-correlated with the density fluctuations. This suggests the dominance of slow-like magnetosonic turbulence \citep[e.g.,][]{S2006,H2015, K2012}.

This is further demonstrated by quantifying separately the contribution of incompressible and compressible fluctuations using the energy cascade rate. In Fig.~\ref{fig:1}(c), we see that the $|\varepsilon_\text{IMHD}|$ and $|\varepsilon_\text{CMHD}|$ almost exactly superimpose to each other at all scales. However, in Fig.~\ref{fig:2}(c) we observe a larger value of $|\varepsilon_\text{CMHD}|$ compared to $|\varepsilon_\text{IMHD}|$. Typically, this increase is due to the compressible flux terms, which include density and internal energy fluctuations \citep[see,][]{A2017b,A2018b}. While at MHD scales $|\varepsilon_\text{CHall}|$ and $|\varepsilon_\text{IHall}|$ in Fig.~\ref{fig:1}(a) behave similarly, a major difference is observed at sub-ion scales: $|\varepsilon_\text{CHall}|$ is increasingly larger than $|\varepsilon_\text{IHall}|$ as one moves towards small scales. This is in agreement with the idea that turbulence transitions into KAW turbulence at these scales \citep{C2017}. This property is also observed for the case with higher density fluctuations in Fig.~\ref{fig:2} and seems to be a fundamental property of MS turbulence. 

This conclusion is clearly demonstrated in Fig.~\ref{fig:4} that shows the mean ratios $|\varepsilon_\text{CHall}|/|\varepsilon_\text{IHall}|$ as a function of the mean ratios $|\varepsilon_\text{CMHD}|/|\varepsilon_\text{IMHD}|$ obtained for all the 72 events that resulted from our data selection. \ADD{We emphasize that here we considered only cases where the dominant cascade rate components showed an approximately constant (negative or positive) sign for all of the time lags in the MHD and sub-ion ranges to ensure a reliable estimate of its mean values. The mean values at the MHD and sub-ion scales were computed over the time lags $\sim [50$ -- $150]$ s and $\sim [0.5$ -- $5]$ s, respectively.} At MHD scales we observe in  Fig.~\ref{fig:4} that higher density fluctuations lead to increasing $|\varepsilon_\text{CMHD}|$ over $|\varepsilon_\text{IMHD}|$. More importantly, we see that even when $|\varepsilon_\text{CMHD}|/|\varepsilon_\text{IMHD}|\sim 1$, most of those cases  show higher $|\varepsilon_\text{CHall}|$ compared to $|\varepsilon_\text{IHall}|$. For some of them, one order of magnitude difference between the two cascade rates is seen. These statistical results support the idea that even a small level of density fluctuations could amplify the energy cascade rate as it goes into the sub-ion scales, demonstrating the inherently compressible nature of the MS plasma turbulence at those small scales. 

\ADD{Finally, in Fig.~\ref{fig:3} we show the mean ratios in the sub-ion scales $|\varepsilon^\text{non-flux}_\text{CHall}|/|\varepsilon_\text{CHall}|$ as a function of the mean ratios in the MHD scales $|\varepsilon^\text{non-flux}_\text{CMHD}|/|\varepsilon_\text{CMHD}|$ obtained for all the 72 events. Similarly to Figs. \ref{fig:1}-\ref{fig:2}(c), we observe that the non-flux terms (i.e., the source, hybrid and $\beta$-dependent \citep[see,][]{A2017b}) are negligible with respect to the flux terms. These observational results are also corroborated by numerical results previously reported in compressible hydrodynamics and MHD turbulence \citep{A2018b}, and in a recent statistical study of the cascade rate at MHD scales using MMS data in the MS \citep{A2019b}.}

%%%%%%%%%%%%%%%%%%%%%%%%%
{\it Conclusion.}
Understanding sub-ion scale turbulence in space plasmas is a difficult subject because the physics involves many processes that we still do not fully understand. Notably, a fundamental question remains open as to how much energy (stirred up at the large scales) leaks down into the sub-ions, which eventually gets dissipated (by some kinetic processes) into ion and/or electron heating? The subsequent question is how much of that energy comes from the incompressible vs. the compressible components of the turbulence? These are fundamental plasma physics questions that are relevant to other distant astrophysical plasmas, e.g. accretion disks of massive objects \citep{S2019, K2019}. In this study, we provide some answers to these questions using the MMS data in the Earth's MS and a recent exact law for compressible HMHD, by estimating the first compressible energy cascade rate at sub-ion scales in a collisionless plasma and demonstrating the leading order played by density fluctuations at those scales. The question as to which kinetic processes dissipates energy into particle heating cannot be addressed by our fluid models. Future studies should tackle this problem and the possible role that kinetic instabilities can play in the cascade at sub-ion scales. 

\section*{Acknowledgments}

N.A.~is supported by a DIM-ACAV post-doctoral fellowship. The authors acknowledge the MMS team for producing the data, which were obtained from the MMS Science Data Center (\url{https://lasp.colorado.edu/mms/sdc/}). The authors acknowledge financial support from CNRS/CONICET Laboratoire International Associ\'e (LIA) MAGNETO.

%N.A.~is members of the {\it Carrera del Investigador Cient\'ifico} of CONICET.


\end{document}